\documentclass[letter]{aa} 
\usepackage{graphicx}
\usepackage{txfonts}
\usepackage[colorlinks=true,citecolor=blue]{hyperref}
\usepackage{graphicx}
\usepackage{times}
\usepackage{natbib}
\usepackage{color}
\usepackage{multirow}
\usepackage{todonotes}
\usepackage{appendix}
\usepackage{array}
\usepackage{verbatim}

\newcommand{\enzo}{\it{\small ENZO}}

\begin{document} 

\title{Exploring the origins of mega radio halos}

\author{L. Beduzzi\inst{1} \fnmsep\thanks{\email{luca.beduzzi@studenti.unipd.it}}
          \and
          F. Vazza\inst{2,3,4}
          \and
          G. Brunetti\inst{3} 
          \and
          V. Cuciti\inst{2,3}
          \and
          D. Wittor\inst{4}
          \and 
          E. M. Corsini \inst{1}
          }
\institute{Dipartimento di Fisica e Astronomia "Galileo Galilei", Università di Padova,
Vicolo dell’Osservatorio 3, 35122 Padova, Italy
    \and 
   Dipartimento di Fisica e Astronomia, Università di Bologna, Via Gobetti 92/3, 40121 Bologna, Italy
         \and
            Istituto di Radio Astronomia, INAF, Via Gobetti 101, 40121 Bologna, Italy
         \and
             Hamburger Sternwarte, University of Hamburg, Gojenbergsweg 112, 21029 Hamburg, Germany
        \and Osservatorio Astronomico di Padova, INAF, Vicolo Dell'Osservatorio 5, 35122, Padova, Italy
        }

   \date{Received 1 May 2023 / Accepted 25 August 2023} 


\abstract    
{We present a first attempt to investigate the origin of radio-emitting  electrons in the newly discovered class of mega radio halos in clusters of galaxies. We study the evolution of relativistic electrons accreted by the external regions of a simulated cluster of galaxies at high resolution, including the effect of radiative losses and turbulent reacceleration acting on relativistic electrons. We conclude that turbulent reacceleration induced by structure formation, if sufficiently prolonged, has the potential to produce a large reservoir of radio-emitting electrons in the large regions illuminated by mega radio halos observed by LOFAR.}


\maketitle

\section{Introduction} \label{sec:introduction}
The unprecedented sensitivity to large-scale diffuse emission provided by the Low Frequency Array (LOFAR; \citealt{2013A&A...556A...2V}) has uncovered extended radio emission at the extreme periphery of clusters of galaxies and between massive pairs \citep[e.g.][]{2019SSRv..215...16V,2019Sci...364..981G,2020MNRAS.499L..11B,2021A&A...648A..11O,2021A&A...654A..68H,bonafede21}, including spectacular and volume-filling radio emission structures up to almost the virial radius in some cases \citep{ 2022SciA....8.7623B}.

In particular, \citet{2022Natur.609..911C} reported the discovery of a new class of large diffuse radio sources outside the central regions of clusters of galaxies, well beyond the volume of `classical' radio halos (classicalRHs), and dubbed mega radio halos (megaRHs). MegaRHs show a rather flat synchrotron brightness profile extending beyond the scale of the classicalRH emission detected in the same clusters. 
Limited to the small statistics of detected objects, the transition from the classicalRHs to megaRHs occurs at a radius that is about one-half of $R_{\rm 500}$ from the cluster centre, and the average emissivity of the megaRHs is about $ 20-25$ times lower than the emissivity of classicalRHs.  For two clusters, the combination of 50 and 140 MHz LOFAR observations allowed  the synchrotron spectrum of the megaRHs to be characterised as a relatively steep spectrum with a spectral index $\alpha\sim-1.6$.

MegaRHs allow us to probe the outskirts of galaxy clusters in an approximately $30$ times larger volume than what has been achieved for decades, and also to study particle acceleration and magnetic field amplification in the complex regime of very weakly collisional plasmas 
\citep[e.g.][]{bj14,2016PNAS..113.3950R, PhysRevLett.131.055201,zhou2023magnetogenesis, 2018ApJ...863L..25S,2022hxga.book...56K}.
The observed steep spectrum in megaRHs would favour a scenario where relativistic electrons are reaccelerated by second-order Fermi mechanisms in a turbulent intraclustesr medium  \citep[ICM; e.g.][]{gb01,2003ApJ...584..190F,cassano05,2016MNRAS.458.2584B}.
However, the discontinuity that is observed in radio brightness profiles of classicalRHs and megaRHs \citep{2022Natur.609..911C} suggests that the latter trace a turbulent component that is different from that of classicalRHs. In fact, numerical simulations show the presence of a baseline level large-scale turbulent component at these cluster radii \citep[e.g.][]{lau09,va11turbo} driven by the continuous accretion of matter onto the cluster and providing significant non-hydrostatic pressure support \citep[e.g.][]{2014ApJ...782..107N,2020MNRAS.495..864A,2023arXiv230315102D}.


In this work, we present the first attempt to evaluate the life cycle of relativistic electrons in the outskirts of galaxy clusters under realistic conditions, with an ad hoc cosmological simulation and methods (Sect. \ref{sec:methods}). We use this approach to estimate the circumstance under which the observed volume of megaRHs can be filled with radio-emitting electrons (Sect. \ref{sec:results}). Our tentative conclusion, rather insensitive to the unconstrained initial distribution of relativistic electrons before the formation of the cluster, is 
that the relativistic particles transported in large volumes may experience second-order Fermi reacceleration on longer spatial and temporal timescales than usually studied for classicalRHs, and may support the emission observed in megaRHs (Sect. \ref{sec:conclusions}).

\begin{figure*}[h!]
    \centering
    \includegraphics[width=0.99\textwidth]{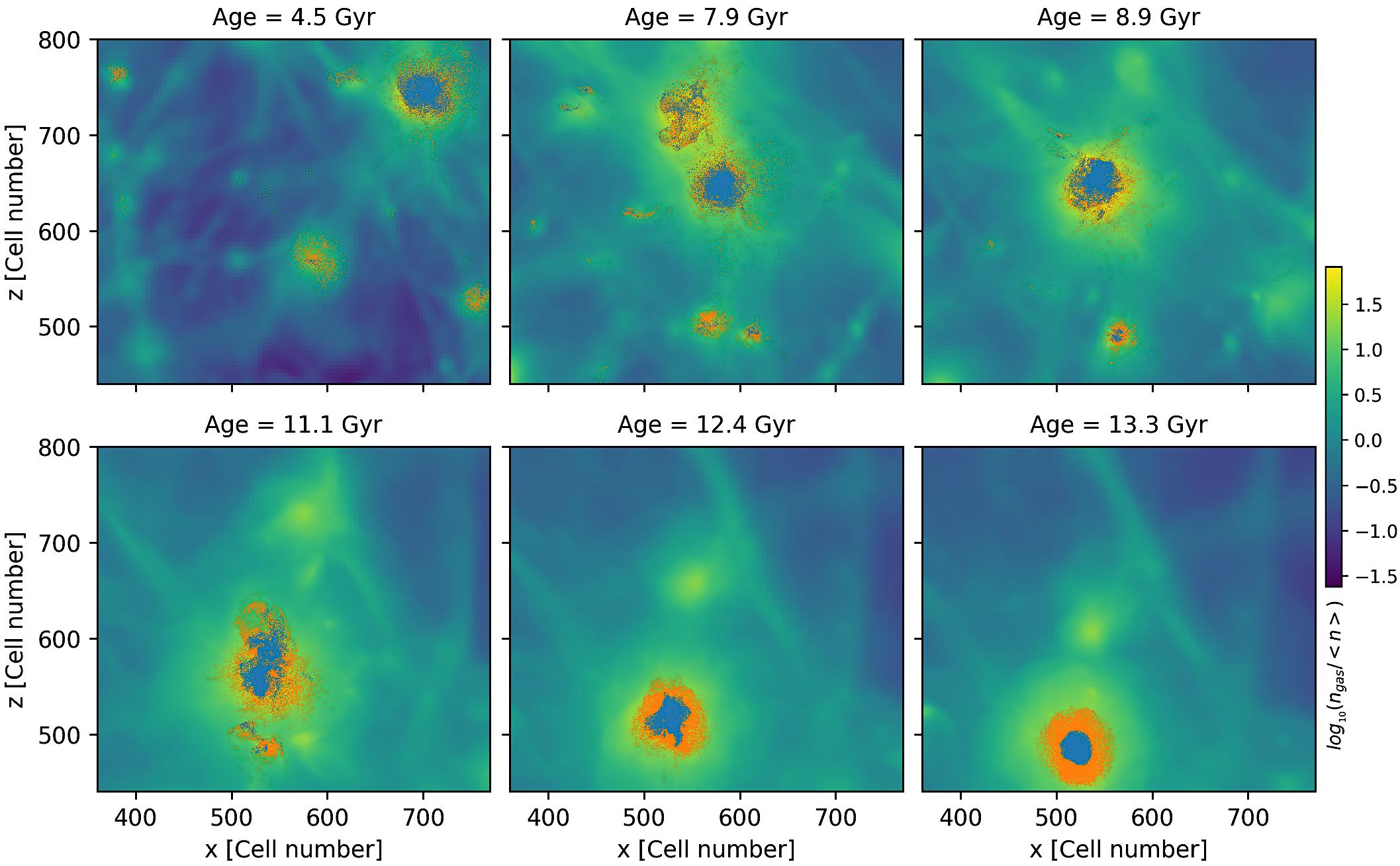}
    \caption{Colour map showing the evolution of the projected comoving gas density (normalised to the total mean matter density) averaged along the full line of sight of the simulation for six different epochs. The points show the projected distribution of the tracers used in our analysis, which are colour coded in orange (for tracers ending up in the megaRH region) or blue (for tracers ending up in the classicalRH region).}    \label{fig:pos_snaps}
\end{figure*}

\begin{figure*}[h!]
    \centering
    \includegraphics[width=0.99\textwidth]{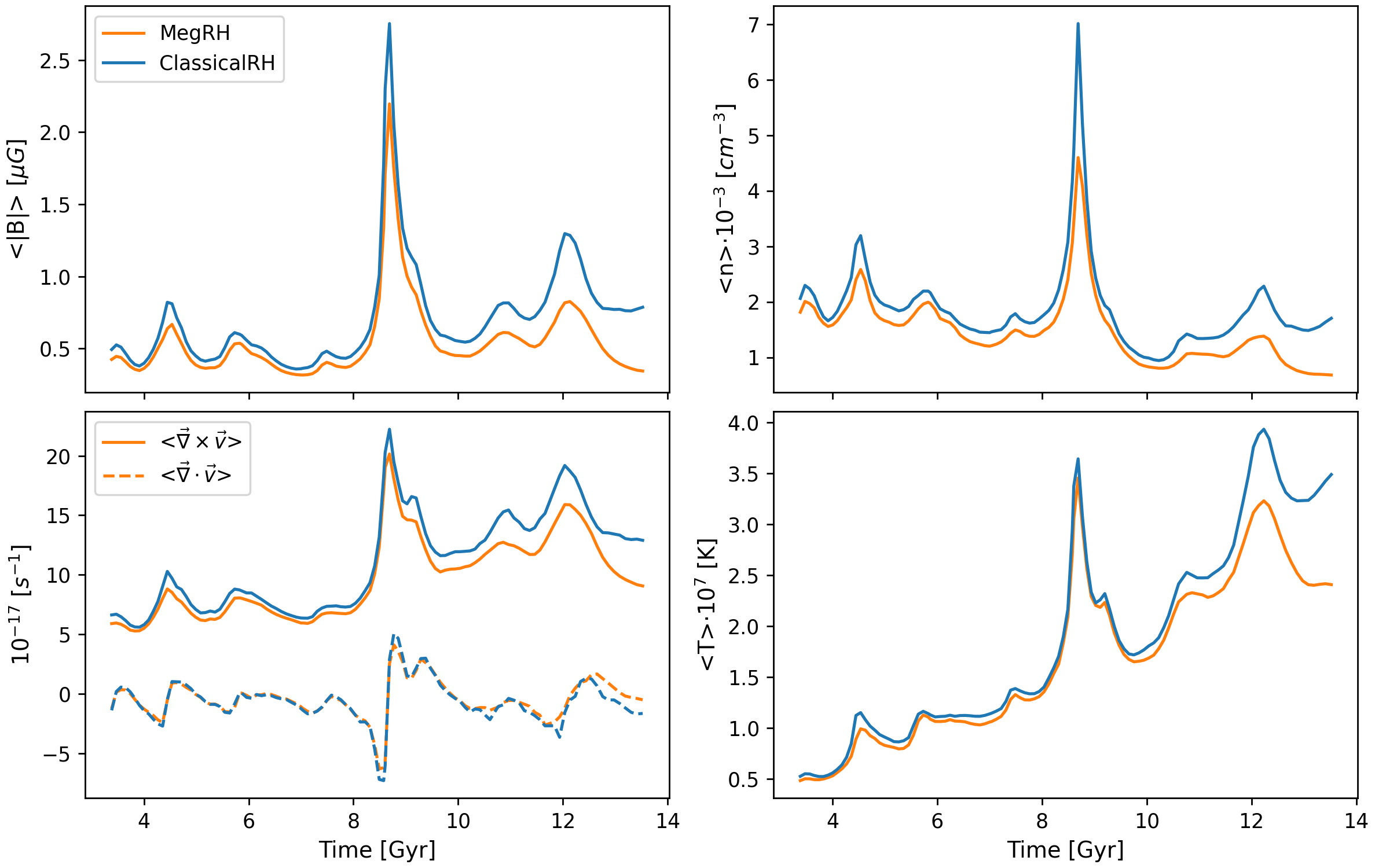}
    \caption{Evolution of mean values of proper magnetic field strength, gas density, vorticity, divergence, and temperature probed by tracers ending up either in the classicalRH (blue) or in the megaRH (orange).}
    \label{fig:trends}
\end{figure*}

\begin{figure}[h!]
    \centering
    \includegraphics[width=0.49\textwidth]{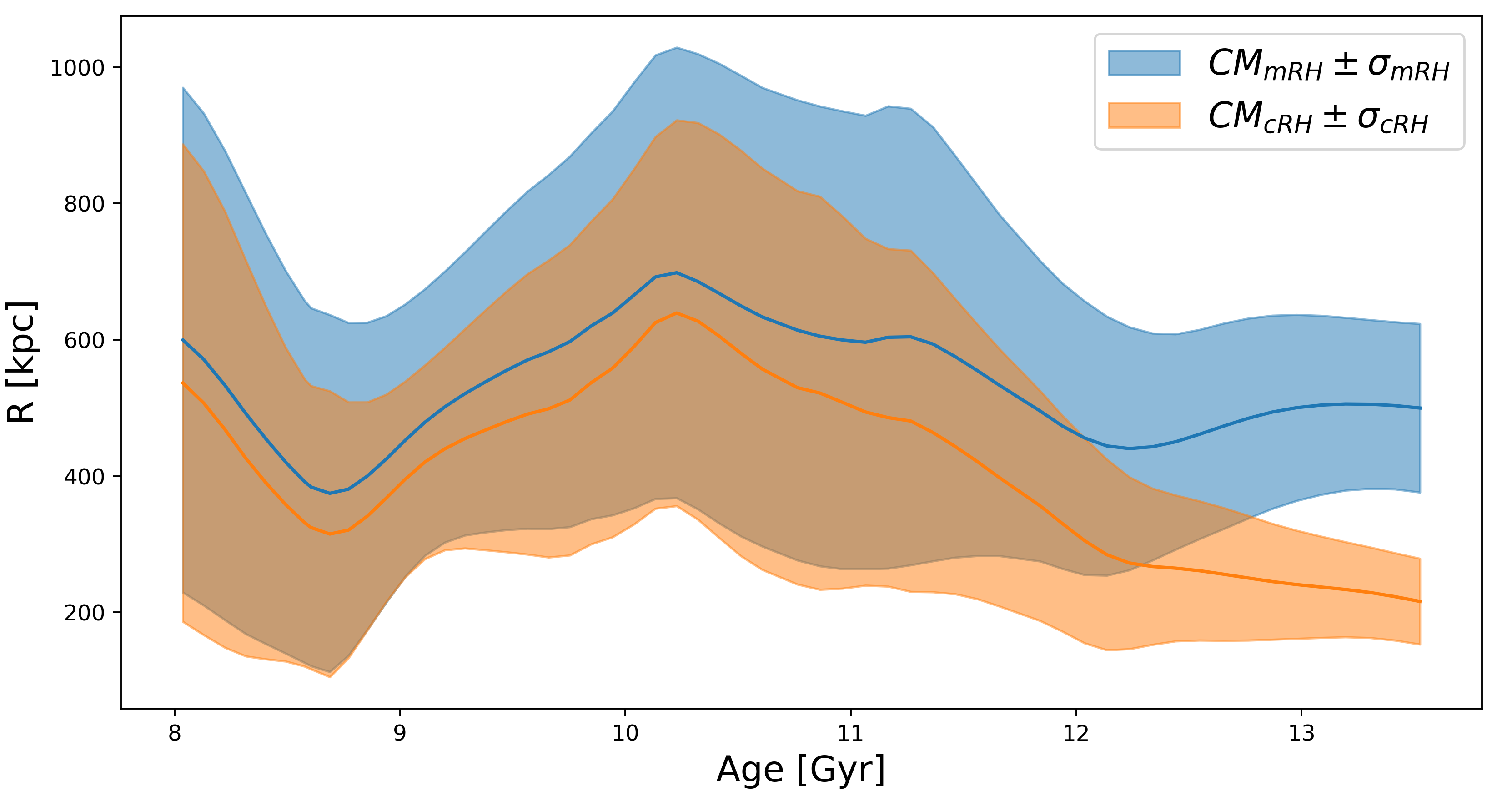}
    \caption{Evolution of the mean distance of our tracers (ending up either in the classicalRH or in the megaRH region) with respect to the moving centre of mass of the cluster as a function of redshift.}
    \label{fig:distance}
\end{figure}

\begin{figure*}[h!]
    \centering
    \includegraphics[width=0.99\textwidth]{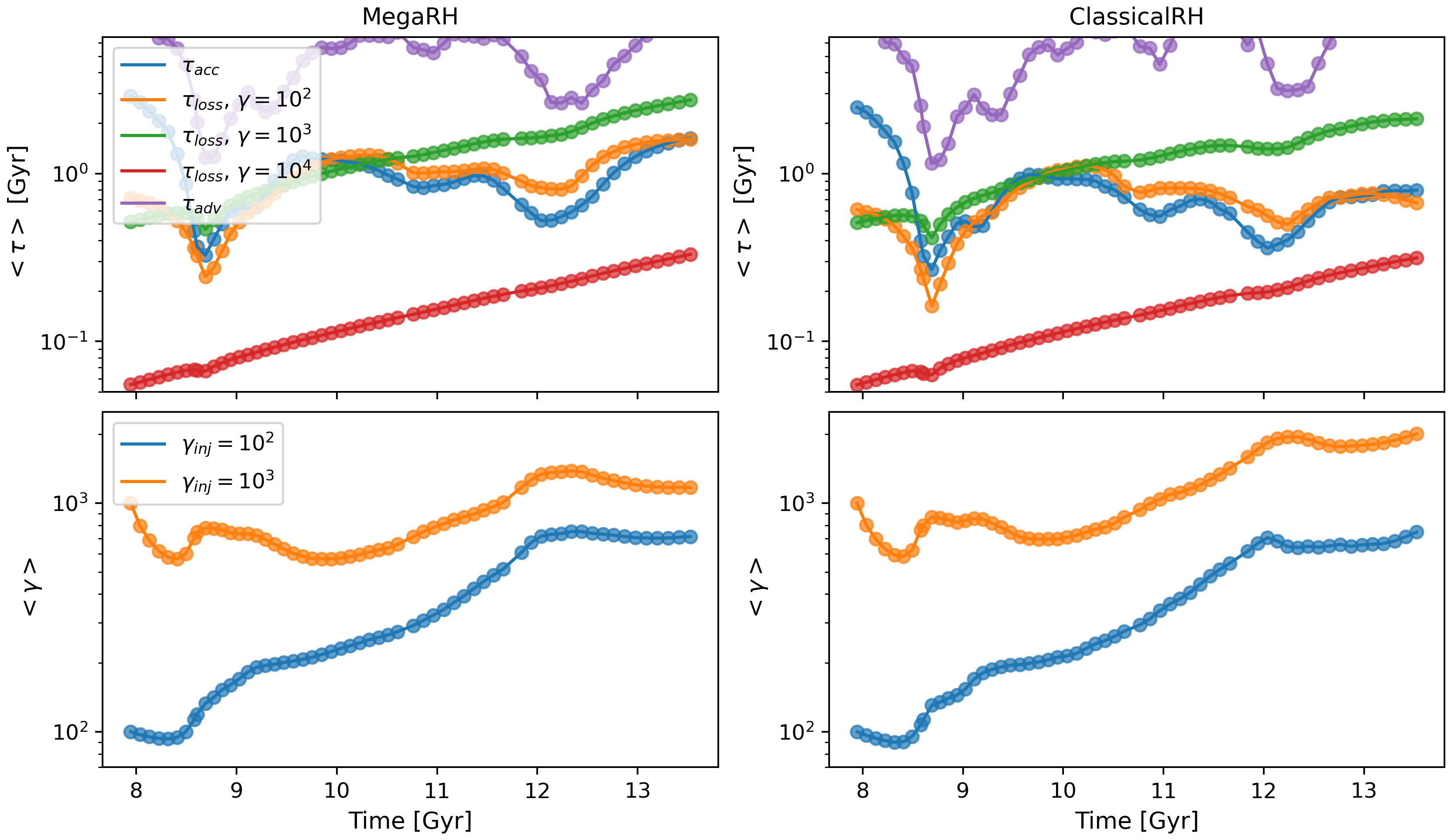}
    \caption{Evolution of the median lifetime of CRe and the evolution of their median Lorentz factor for the different families of tracers. Top panels: Evolution of the mean values of the loss or reacceleration timescales for different electron energies, for tracers ending in the classicalRH or megaRH regions (left). Lower panels: Evolution of the mean $\gamma$ of electrons starting from  $\gamma_{\rm inj}=100$ or  $1000$, and ending up  either in the megaRH (left) or classicalRH (right) region.} 
    \label{fig:timescales}
\end{figure*}

\section{Methods and simulations} \label{sec:methods}
\label{subsec:sim}
We produced a cosmological adaptive-mesh refinement {\enzo} simulation of a cluster of galaxies using self-gravity for ordinary and dark matter, radiative equilibrium cooling, and no other galaxy formation-related physics (e.g. star formation or feedback from supernovae).  Magnetic fields are evolved assuming  ideal magnetohydrodynamics  (MHD) using the hyperbolic cleaning method \citep[][]{enzo14} (see \citealt{dom19} for previous tests and results with a similar numerical setup). We started from a simple uniform magnetic field seed of $B_0=0.4  \ \mathrm{nG}$ (comoving) in each direction at $z=40$. 

The simulated cluster has a total (ordinary+dark) matter mass of $M_{\rm 100}=3.8 \cdot 10^{14} ~M_{\odot}$ and a virial radius of $R_{\rm 100}=1.52~ \rm Mpc$ at $z=0$.
It is the most massive object to form in a total volume of 
(100 Mpc$/h$)$^3$, sampled with a uniform  root grid of $128^3$ cells and dark matter particles (with a mass resolution of $m_{DM0}=3 \cdot 10^{10} M_{\odot}$), and further refined with four additional levels of $\times 2$ refinement in spatial resolution each (and $\times 2^3$ refinement in mass resolution of the dark matter component at each refinement step) with nested regions of decreasing size.
The inner (15.6 Mpc)$^3$ volume, where the cluster forms, is resolved with a maximum dark matter mass resolution of $m_{DM}=7.3 \cdot 10^6 M_{\odot}$ (i.e. $m_{DM0}/8^4$), and with a uniform spatial resolution of $70$ kpc/cell. We also allowed for two extra levels of adaptive mesh refinement within this innermost region in order to best model turbulence by refining on local gas or dark matter overdensities up to a final maximum resolution of $\approx 12.2 \rm ~kpc/h= 18.0$~kpc (comoving). This resulted in central magnetic field values of $\sim 0.5 ~\rm \mu G$ in the cluster centre, which signals an excessively low amplification efficiency for the small-scale dynamo captured at this finite spatial resolution \citep[e.g.][]{2016ApJ...817..127B}.   In any case, for all our calculations  we used a rescaled magnetic field intensity in order to take into account the unresolved small-scale amplification. 
The realistic picture of magnetic field amplification in the collisionless plasma of clusters of galaxies is very likely to be much more complex than the MHD approximation used here \citep[e.g.][]{2004ApJ...612..276S,2016PNAS..113.3950R}, which nevertheless yields realistic values of magnetic fields for our regions of interest.
In detail, following \citet{bv20}, we renormalised in the value of $B$ to be assigned to each tracer particle  post-processing under the realistic expectation of efficient small-scale dynamo amplification. More specifically, we assumed that after the turbulent kinetic energy cascade reaches dissipation scales, a fixed fraction $\eta_B$ ($=3\%$) of the energy flux of turbulence is dissipated into the amplification of magnetic fields, as in \citet[][]{bm16}.  Therefore, the magnetic field assigned to each tracer follows from $B_{\rm turb}^2/8\pi \sim \eta_B F_{\rm turb} \tau \sim {1 \over 2} \eta_B \rho \delta V_{\rm turb}^2$, where $\tau= L/\delta V_{\rm turb}$ is the turnover time,  and $\delta V_{\rm turb}$ is the turbulent velocity, which is estimated for each tracer as $\delta V_{\rm turb}= |\nabla \times \vec{v}| L$, that is, the gas vorticity measured across a stencil $L$ ($=54 \rm ~kpc$ in our case, considering a stencil of three cells), which is is our best estimate of the solenoidal turbulent velocity.  Whenever $B_{\rm turb} \geq B$ (where $B$ is the magnetic field recorded by a tracer), we replaced the value of the actual numerical simulation with $B_{\rm turb}$, and used this to compute radiative losses and the Fermi II reacceleration term (see following section). To give some reference numbers, our tests in \citet{va21jets} showed that the median values of the actual magnetic field produced by the simulation and the one computed a posteriori assuming dynamo amplification differ by a factor of approximately two at most for tracers within the volume of clusters, and that in no case would the rescaled magnetic field lead to an energy comparable to the thermal gas pressure.

\begin{figure*}
    \centering
             \includegraphics[width=0.99\textwidth]{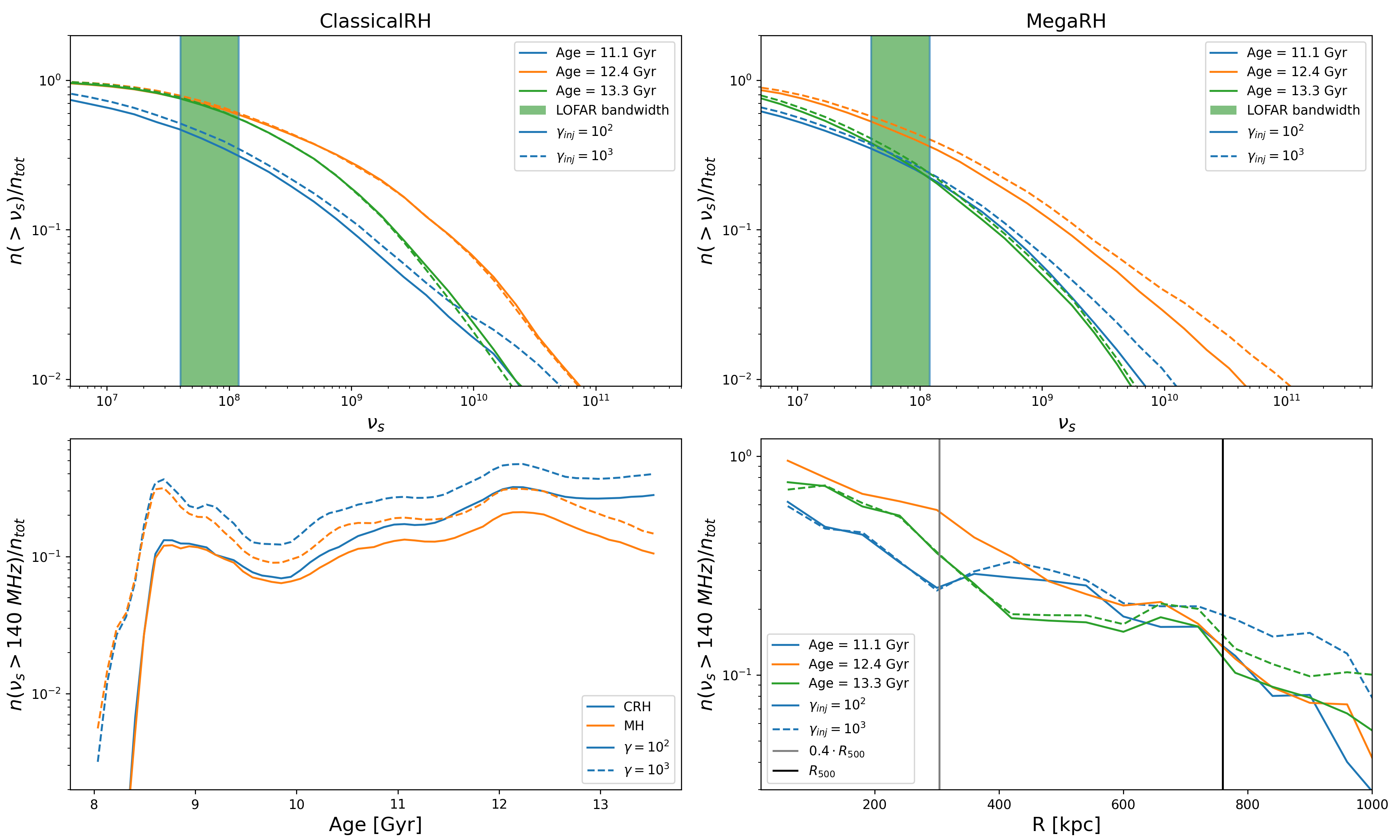}
    \caption{Fraction of visible electrons with respect the frequency, the time and the distance from the center of the cluster. Top panels: Distribution of emission frequencies for the classicalRHs and megaRHs for three epochs (11.1, 12.4, and 13.3 Gyr) and  $\gamma_{\rm inj}=10^2$ or $=10^3$. The vertical green areas show the frequency range covered by LOFAR LBA and HBA. Bottom left panel: Evolution of the fraction of tracers whose predicted emission frequency $\nu_{\rm c}$ is larger than the central LOFAR HBA frequency, for the same models. Bottom right panel: Radial profile of the fraction of tracers whose emission frequency $\nu_ {\rm c}$ is  larger than the central LOFAR HBA frequency, for the same epochs as those shown above.} 
    \label{fig:freq}
\end{figure*}

\subsection{Energy evolution of relativistic electrons}
\label{subsec:times}
In post-processing, we  injected and propagated approximately $10^5$ passive tracers, which allowed us to track the Lagrangian history of gas matter in this system with the  CRaTer code \citep[][]{wi17}. The distribution of particles was created by sampling a fixed gas mass resolution of $\approx 5 \cdot 10^8 M_{\odot}$ in the 3Dl distribution of gas produced by the MHD simulations, and by evolving the positions of particles  forward in time with a simple time integrator $\vec{r}(t+dt) = \vec{r}(t) + \vec{v_{\rm gas}} dt $ after interpolating the 3D velocity field with a  cloud-in-cell spatial interpolation method, as explained in \citet{wi17}. We used the time resolution of our finest snapshot recovery scheme, $dt \approx 90 \rm ~Myr$ (nearly constant), which gave us 103 snapshots to analyse. Figure \ref{fig:pos_snaps} shows six epochs in our simulation, and the evolving positions of the tracers. 
Each tracer records the time evolution of various physical quantities of interest (gas density, velocity, divergence, vorticity, temperature, and magnetic field intensity) and is meant to track the spatial propagation of families of relativistic electrons frozen onto the gas by the tangled magnetic field and to neglect the small effect of diffusion for cosmic ray electrons (CRe) in this energy range \citep[e.g.][]{bj14}. 

As opposed to the case where more computationally expensive approaches are taken, in this paper we are interested in the energy evolution of electrons rather than in their exact spectra, and therefore we do not make use of Fokker-Planck methods and simply assume that each tracer samples the energy evolution of a specific energy of electrons. We use the `ultra-relativistic' approximation $\gamma \approx E/mc^2$ and computed the combination of loss and gain terms as
$    \dot{\gamma} \approx  |{\gamma}/{\tau_{\rm rad}}| + |{\gamma}/{\tau_{\rm c}}| +
    {\gamma}/{\tau_{\rm adv}} - |{\gamma}/{\tau_{\rm acc}}|$, 
where $\tau_{\rm rad}$, $\tau_{\rm c}$, and $\tau_{\rm adv}$ are respectively the loss timescales for the radiative, Coulomb, and expansion (compression) processes, while 
$\tau_{\rm acc}$ represents the acceleration timescale due to turbulent reacceleration. All loss terms are as in \citet{bj14}.
The median gas velocity dispersion of our tracers is $\sigma_v  \sim 500 \rm ~km/s$, and is up to $\sim 700 \rm km/s$ during mergers.

Turbulence may (re)accelerate relativistic electrons in many ways through resonant and non-resonant mechanisms \citep[e.g.][]{1987A&A...182...21S,longair,2006ApJ...638..811C,bl07,2012SSRv..173..535P,bj14,2020LRCA....6....1M,2020MNRAS.499.4972L}. Here, we follow  \citet[][]{2016MNRAS.458.2584B}, and assume that relativistic particles are reaccelerated in a systematic way in reconnecting and magnetic-dynamo regions, and that they undergo a stochastic Fermi II process  overall  on longer timescales, diffusing across incompressible super-Alfvenic turbulence. This mechanism was shown to be a promising candidate to explain the production of diffuse radio emission at the periphery of clusters of galaxies \citep[e.g.][]{bv20,bonafede21,2022SciA....8.7623B}.

The acceleration timescale for the turbulent reacceleration process in \citet[][]{2016MNRAS.458.2584B} is: 
\begin{equation}
    \tau_{\rm acc} =  125  \rm ~Gyr \rm  \frac{L[\rm kpc]/(500) ~ B[\rm \mu G]}{\sqrt{n[\rm cm^{-3}]/10^{-3}} ({\delta V_{\rm turb}[\rm cm/s]/10^7})^3} ,
    \label{eq:ASA}
\end{equation}

\noindent with $\delta V_{\rm turb}$ and $L$ as defined above. 
Based on previous work with identical simulations of cluster physics, we can  reasonably approximate that the turbulent spectrum of velocity fluctuations follows a power-law energy distribution, with a well-defined inertial range in which the Kolmogorov scaling is closely followed \citep[e.g.][]{va11turbo,2022A&A...658A.149S}, both for the compressive and for the solenoidal components of the flow \citep[][]{va17turb}.
In this case, Eq.1 does not depend on the  exact choice of the scale $L$ where we measure vorticity, because the turbulent energy flux ($F_{\rm turb} \propto \delta V_{\rm turb}^3/L$) is constant within the  inertial range of turbulence \citep[see a longer discussion in][]{va21jets}.

Other mechanisms can contribute to the reacceleration of particles in the turbulent ICM. A popular mechanism that is commonly assumed to model turbulent reacceleration in galaxy clusters is based on transit time damping (TTD), which itself is based on resonant acceleration with compressive fast modes \citep[e.g.][]{bl07,2017MNRAS.465.4800P}. This mechanism is expected to generate acceleration that are faster than those of the slow solenoidal mode 
for (compressive) turbulent Mach numbers $\mathcal{M}_c^2 \geq 0.3-0.4$ \citep[e.g.][]{2016PPCF...58a4011B,bv20}. As a consequence, in the presence of a significant component of compressive turbulence in the ICM, the coexistence of our baseline reacceleration and the TTD mechanism is likely to imply that electrons may end up in the megaRH regions after having experienced a substantially higher reacceleration.

\section{Results} \label{sec:results}
The aim of this work is to test whether the turbulence measured in the external regions of the simulated cluster is sufficient to generate a population of radio-emitting electrons filling a large intracluster volume, which is  similar to what has been observed in megaRHs by LOFAR. Furthermore, by sampling the cluster volume with tracers, we want to evaluate whether the evolution and effective lifetime of relativistic electrons in the region of classicalRHs are similar to those of electrons found in the periphery. 

First, we measure the typical evolution of the physical conditions of gas matter accreted onto classicaRH and megaRH regions at the end of the simulation. We can do this in a straightforward way because our tracer approach allows us to study the Lagrangian evolution over time of the gas matter found anywhere in the simulation at a given time.
Based on the radio surface brightness profiles of the four clusters studied in \citet{2022Natur.609..911C}, we define the classicalRH region as $0 \leq r \leq 0.4R_{500}$, where $r$ is the radial distance from the centre of mass of our cluster at any given epoch, while we define the megaRH region as the region in the $0.4R_{500} \leq r \leq R_{500}$ range. 

Figure \ref{fig:trends} gives the 
time evolution of the median values of magnetic field, gas density, curl, and divergence of the gas velocity field and of temperature for tracers found in the two regions towards the end of our simulation.   This clearly shows that the (thermo)dynamical history of gas matter ending up in either the classicalRH or megaRH region is very similar. It should be noted that, although the average volume-weighted gas density in megaRH is about a factor ten smaller than in classicalRHs, our tracers give a mass-weighted average of all fields that is biased towards the clumpiest end of the gas distribution.
The bulk of gas accreted by the megaRH regions has already crossed the cluster centre once, and has been dispersed to larger radii only towards the end of the evolution.
This is confirmed by Figure \ref{fig:distance}, which shows the mean distance of our tracers (selected in the two regions) from the moving centre of mass of the cluster as a function of time: only in the last $\sim 1.5 \rm ~Gyr$ of evolution does the average distance of the two populations become significantly different, while it was nearly the same before then, suggesting once more that the gas ending up in the two regions shared a similar history for all previous evolutionary stages.
The relativistic electrons carried by this accreted gas have already been subject to significant turbulent reacceleration (and compression) once in the past, while in the final part of their evolution they age in a less dense and magnetised environment than classicalRHs, which makes their loss timescale slightly longer (although inverse Compton losses are always present). 

This makes the gas matter in the megaRH region suitable for the reacceleration of relativisic electrons up to and beyond $\gamma \sim 10^3$ (on average), allowing a fraction of them to become radio emitting, similar to the matter ending up in the classicalRH region. 
This is shown in Figure \ref{fig:timescales}, which presents the evolution of the median timescales for acceleration and loss processes for electrons with fixed energies: $\gamma=10^2$, $10^3$, or $10^4$. 
The same figure also shows the predicted 
evolution of $\gamma(t)$ for our population of tracers, which was obtained by directly integrating $\gamma(t) = \gamma_{\rm inj} + \int (d\gamma/dt) dt$ in time, where the loss and acceleration terms are computed as in Sect. 2.1 and for which we tested two different initial values of the seed electrons, $\gamma_{\rm inj}=100-1000$.
Even in the (very conservative) scenario in which all seed relativistic electrons have $\gamma_{\rm inj}=10^2$  before their later accretion onto the main cluster, the combination of compression and turbulent reacceleration in the subsequent dynamics allows them to reach $\gamma \sim 500-700$, on average, when the cluster merger occurs.
The fact that the electrons accreted in these regions can keep their initial energy (and even increase it by a factor of a few) instead of losing it within their respective cooling time is a fundamental finding of this work (see also Fig. 6 and Sect. \ref{sec:conclusions} below).
For example, taking $\gamma=10^2$ electrons  in Fig.\ref{fig:timescales}  as a reference, it is clear that while their loss timescale at the epoch of 8 Gyr ($z \approx 0.6$) is of $\tau_{\rm loss} \sim 0.7 \rm ~Gyr$, their energy is not lost after $\tau_{\rm loss}$, but rather continues to increase on average for the following 5 Gyr. This happens because the mild level of turbulent reacceleration experienced by electrons accreted onto the megaRH region (or additionally, adiabatic compression in the case of classicalRH) can increase their energy on a similar timescale. 
Therefore, in this system, the turbulent reacceleration induced by the merger (epoch of $\approx 12 \rm ~Gyr$) acts on an already `preaccelerated' pool of relativistic electrons. 

Next, we estimate the fraction of the accreted gas matter that may produce radio emission at  LOFAR frequencies given its local plasma conditions and its past sequence of loss or gain mechanisms. 
In the turbulent reacceleration scenario, there is a critical Lorentz factor $\gamma_{\rm c}$ at which the total cooling timescale $t_{\rm cool}$ becomes comparable to the acceleration timescale $t_{\rm acc}$.  Correspondingly, the emission radio spectrum will steepen at a frequency of $\nu_{\rm c}=\xi \nu_{\rm b}$, where $\nu_{\rm b}$ is the break frequency and $\xi\sim 6-8$; the former can be estimated from the maximum value of the Lorentz factor of the electrons associated with each tracer, that is, $\nu_{\rm b}\approx 4.6 \langle \gamma \rangle^2 \cdot (B/\rm \mu G) \rm [Hz](1+z)^{-1}$ \citep[][]{2010A&A...509A..68C}. This allows us to translate the $\gamma(t)$  of our tracers measured onboard
as a function of time into an estimate of the maximum observable synchrotron emission frequency. As a first-order approximation, the fraction of tracers for which $\nu_{\rm c}$ is larger than the LOFAR High Band Antenna (HBA, 140~MHz) or LOFAR Low Band Antenna (LBA, 50~MHz) central observing frequencies can tell us the fraction of the simulated volume that will potentially
become radio observable as a result of the turbulent reacceleration process. These results can also be used in a  straightforward manner to predict the observability of megaRH in other radio telescopes routinely used to study classicalRH (e.g. with the uGMRT, e.g. \citealt{2020IAUS..342...37K}, the MWA, e.g. \citealt{2021PASA...38...53D}, MeerKAT e.g. \citealt{2022A&A...657A..56K}) as well as with the future Square Kilometer Array \citep[e.g.][]{2013A&A...554A.102G,2015aska.confE..92J}.

The upper panels of Figure \ref{fig:freq} show the cumulative distribution of the emission frequencies from all our tracers at three different epochs (just before, during, and after the last merger of the cluster) and for different initial values of the initial energy of electrons and separately for those located in the classicalRHs  and in the megaRHs. The lower panels show the time evolution of the fraction of  tracers in which the computed emission frequency $\nu_c$ is larger than the LOFAR HBA observing frequency (150 MHz), as well as the radial profile of the same fraction for the same three reference epochs as in the upper panels. In both regions, a large fraction of the tracers gets accelerated beyond the 50-140 MHz range of LOFAR: $\approx 31\%-79\%$ of the volume of the classicalRH and $\approx 22\%-57\%$  of the volume of the megaRH should emit detectable radio emission in the three investigated epochs. 
The above fractions are only slightly changed when using $\gamma_{\rm inj}=10^2$ instead of $\gamma_{\rm inj}=10^3$. Therefore, during mergers, the average energy of particles is increased by turbulent reacceleration to a level at which there is approximate balance between gain and loss terms (from synchrotron and inverse Compton emission), and the memory on the initial energy of the population of seed electrons is nearly lost.  The full time evolution shows that, even if mergers obviously enhance the fraction of radio-emitting particles, the continuous turbulent reacceleration experienced by the accreted gas always keeps a significant fraction of them near to LOFAR frequencies. 
Intriguingly, the radial decline of the profile qualitatively matches the observed drop in the radial profile of the surface brightness observed in real megaRHs \citep[][]{2022Natur.609..911C}. However, in the future, we will need to resort to more computationally demanding Fokker Planck methods \citep[e.g.][]{2014MNRAS.443.3564D,2020MNRAS.491..993G,2023MNRAS.519..548B} in order to fully simulate the synchrotron emission spectra of these accelerated cosmic ray electrons.


\section{Discussion and conclusions} \label{sec:conclusions}

The motivation for this work is the recent  discovery of mega radio halos in clusters of galaxies \citep[][]{2022Natur.609..911C}.
We produced a new cosmological simulation of the evolution of a cluster of galaxies to test whether or not the turbulence generated in its outskirts is able to maintain sufficiently energetic ($\sim \rm GeV$) electrons in a fraction of the volume.

We simulated the formation of one cluster of galaxies with the cosmological code {\enzo}  and used Lagrangian tracer particles to follow the energy evolution of families of electrons under the effects of radiative and Colomb losses, adiabiatic changes, and turbulent reacceleration.  Among the several possibilities of particle acceleration in the turbulent ICM, we limited our analysis to the mechanism proposed by  \citet[][]{2016MNRAS.458.2584B}. 

We find that the formation of megaRHs, despite their unprecedented size, can be understood within the framework of the turbulent reacceleration models already used to interpret the classicalRH phenomenology \citep[e.g.][]{bj14}: regardless of the initial energy of cosmic ray electrons, the integrated effect of turbulent reacceleration by cluster-wide turbulent motions is enough to make $50\%$ of our tracers in the megaRH volume able to radiate in the LOFAR band (50-140 MHz). 

Our Lagrangian analysis of the megaRH region suggests a few crucial and overlooked aspects of electron acceleration in these peripheral regions.
First, our tracer analysis allows us to show that most of the matter found in the megaRH comes from the disruption and mixing of gas matter initially located in clumps, which was accreted some gigayears before the final merger by the main cluster, and does not instead come from smooth gas accretions (Fig. 1). 

Connected to this, the dynamical histories of the particles in the classicalRH or in the megaRH are very similar, except in the last 1 Gyr after the latest merger (Fig. 2). Thus a large fraction of the baryons has already crossed the innermost cluster regions at least once, and the relativistic electrons in the megaRH have been processed at least once by the merger.

Finally, our results call for a substantial update of the theoretical picture of the lifetime of relativistic electrons in the ICM. The complex turbulent dynamics therein generates multiple episodes of reacceleration, which overall prolongs the effective lifetime of electrons beyond what was previously thought. This implies that the pool of `fossil' electrons in the ICM may be significantly more energetic than expected based on previous findings.

This is best summarised by Figure \ref{fig:times}, in which we show the computed typical lifetime of CR electrons belonging to the classicalRH and to the megaRH region as a function of their energy and following \citet{bj14}. 
We can define the CRe lifetime $\tau_{\rm CRe}$ as:

\begin{figure}[h!]   
\includegraphics[width=0.49\textwidth]{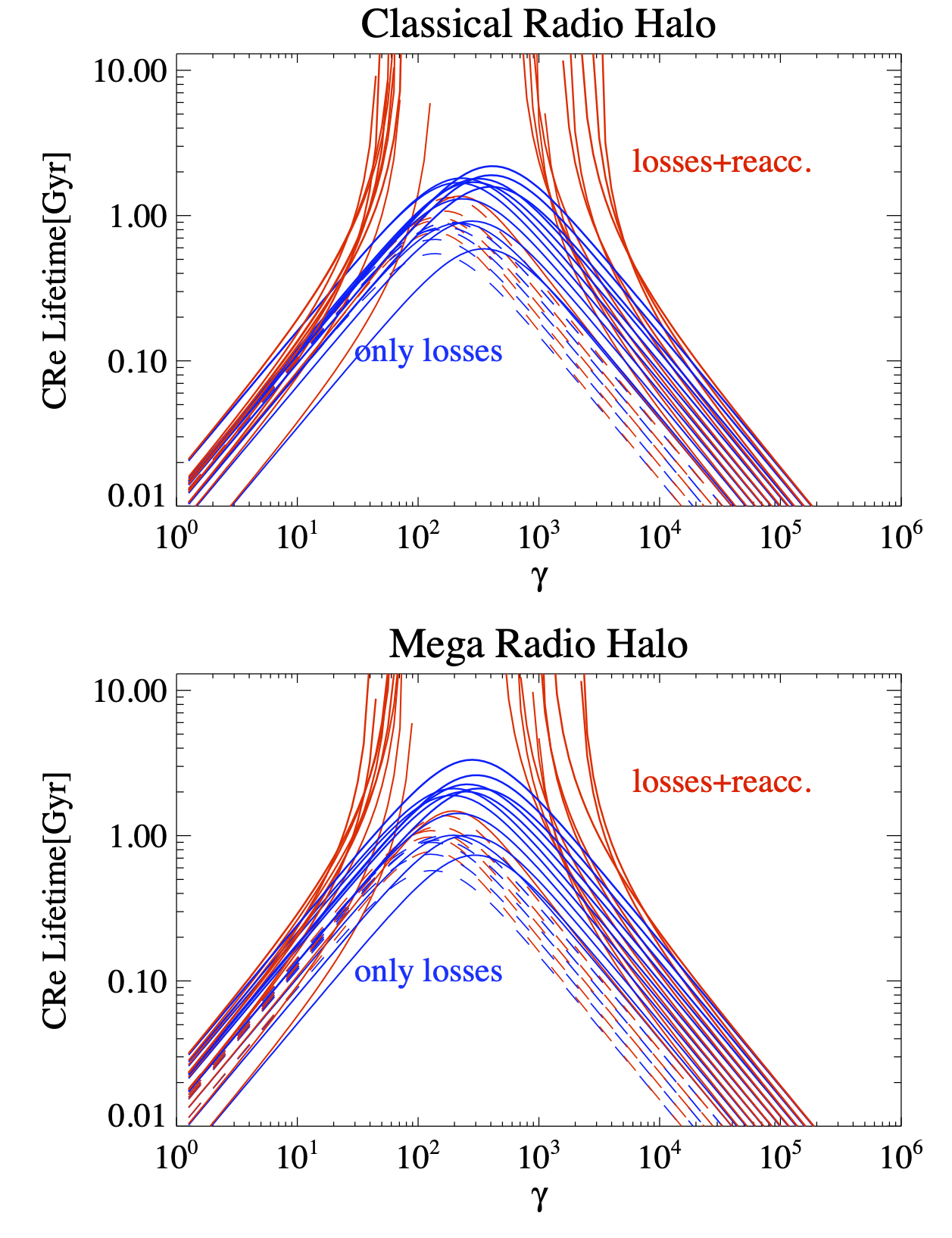}
\includegraphics[width=0.49\textwidth]{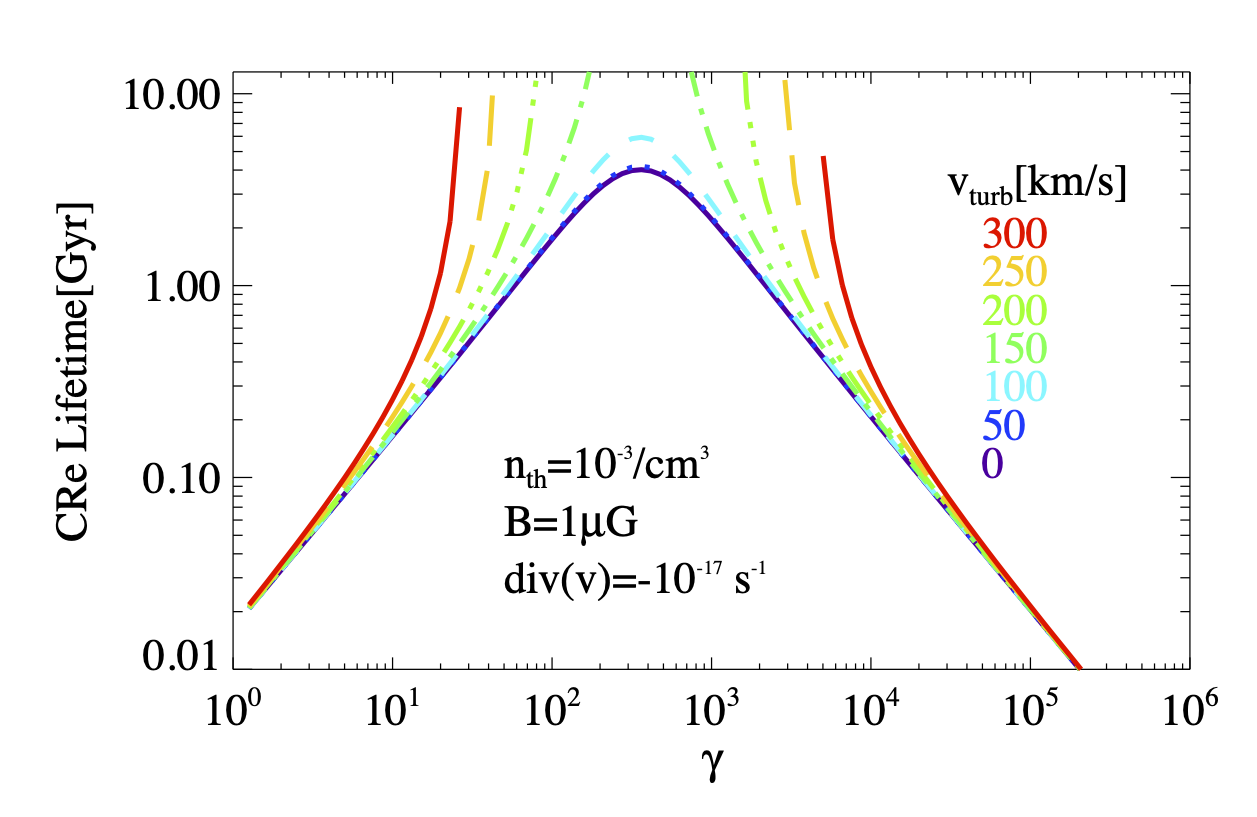}
    \caption{In these figures we report the evolution of a CRe lifetime for different field and time values. First two panels: Lifetime of CRe as a function of their energy, based on loss processes only (blue lines) or also including the balancing effect of turbulent reacceleration and adiabatic compression (red lines) for particles in the classicalRH (top) or in the megaRH (bottom) regions.  We used the (median) physical values of particle density, magnetic field strength, divergence, and vorticity for equally spaced snapshots ($\Delta t\approx 0.7 \rm ~Gyr$) in our simulation, starting from z=1. Dashed lines are for the $z>0.5$ snapshots, with solid lines otherwise. Bottom panel: Lifetime of cosmic ray electrons as a function of their energy for fixed reference values of $B=1 \rm ~\mu G$, $n=10^{-3} \rm ~/cm^{3}$, $\nabla \cdot \vec{v}=10^{-17} \rm ~s^{-1}$, and $z=0.0$ and for increasing values of the solenoidal turbulent velocity.} 
    \label{fig:times}
\end{figure}


\begin{equation}\label{eq:tau_cre}
    \begin{split}
        &\tau_{\rm CRe}[\rm Gyr] = \big|\frac{1}{\tau_{\rm loss}} - \frac{1}{\tau_{\rm gain}}\big |^{-1} = \\
        &=\frac{\tau_{\rm gain}}{|\frac{an_{\rm th}}{\gamma}[1+q~ln (\frac{\gamma}{n_{\rm th}})]+b\gamma[(\frac{B}{3.2 \rm \mu G})^2+(1+z)^4] \tau_{\rm gain}-1|}
    \end{split}
,\end{equation}

in which the thermal gas density $n_{\rm th}$ is measured in units of $\rm part/cm^3$, all times are measured in gigayears, and we used the constant values of $a=38$, $b=41,$ and $q=0.092$. The gain timescale stems from the combination of the turbulent acceleration timescale (Eq. \ref{eq:ASA}) and the advection timescale related to compression and expansion, $  \tau_{\rm adv} = \rm 951 \rm ~Myr/[\nabla \cdot \vec{v}/(10^{-16} \rm s^{-1})]$, so that $\tau_{\rm gain}=(1/\tau_{\rm acc}+1/\tau_{\rm adv})^{-1}$. 
The lines give our estimate of the CRe lifetime $\tau_{\rm CRe}$ based on Eq. \ref{eq:tau_cre} above and using the median values of gas density, magnetic field strength, velocity divergence, and vorticity at each timestep (as in Fig. \ref{fig:trends}), either considering only loss processes (blue lines) or also including the reacceleration by turbulence or gas compression. Here, we consider an evenly spaced sequence of $z<1$ snapshots, in which we additionally marked the $z<0.5$ data  with solid lines, which approximately mark the start of the final merger of the main cluster. 

It should be noted that the timescales related to the spatial diffusion of CRe are instead much larger than all the times in the energy range considered here. For a range of plausible spatial diffusion coefficients of CRe propagating on large scales in the tangled ICM magnetic fields, $D \leq 10^{30}-10^{31} \rm cm^2/s$ \citep[e.g.][and discussion therein]{bj14}, the diffusion timescale over a $L \approx 1 \rm ~Mpc$ scale of megaRH (as well as classicalRH) is longer than both the Hubble time and any of the processes acting on radio-emitting electrons: $\tau_{\rm diff} \sim L^2/(4D) = 30-300 \rm ~Gyr \gg t_H$.

As can be seen in the figure, starting from  approximately $z \sim 0.5$ (i.e. from an epoch of $\sim 9 \rm ~Gyr$),  the effect of turbulent reacceleration (combined with that of gas compression) is able to extend the lifetime of CRe beyond what radiative losses alone will allow. Starting from $z \sim 0.5$, the level of turbulence experienced by tracers is large enough to give CR electrons in the  $\gamma \sim 50 - 2000$ energy range a formally infinite lifetime. Moreover, during the most active stages of the merger, particles up to $\gamma \sim 10^4$ have a significantly longer lifetime than what is allowed by their radiative lifetime, once more confirming that turbulent reacceleration ---also in the case of megaRHs--- is enough to produce a significant reservoir of potentially radio emitting electrons.  Finally, in the last panel of Fig. \ref{fig:times} we study the dependence of the CR electron lifetime as a function of the turbulent velocity (while we fix the other parameters to the typical ICM values of $B=1 \rm ~\mu G$, $n=10^{-3} \rm ~/cm^{3}$, $\nabla \cdot \vec{v}=10^{-17} \rm ~s^{-1}$ and $z=0.0$): for turbulent velocities of $\geq 100 ~\rm km/s$ (within our stencil scale of $54 \rm~ kpc$), the lifetime of $\gamma \sim 200-800$ electrons becomes infinite, while for turbulent velocities of $\geq 250 ~\rm km/s,$  radio emitting electrons ($\gamma \geq 5000$) 
 can also have infinite lifetimes.

In this work, we did not produce a detailed simulation of the possible injection process of the relativistic electrons later used for the production of radio emission, as in other works \citep[e.g.][]{va21jets,va23a}. However, the fact that the bulk of the gas content of megaRHs comes from the accretion of already formed halos at $z=2$ ---as shown by our simulation--- means that it is extremely likely that each of these halos has a significant content of relativistic
electrons, following from the activity of star formation and AGN, which both peak around this epoch. In this sense, our initial energy of $\gamma_{\rm inj} = 10^2$ should be regarded as a very conservative limit on the true energy of the relativistic electrons carried by the accreted halos.


By building on this first positive test of the turbulent reacceleration hypothesis for the origin of megaRHs, in forthcoming works we will perform complete simulations, both of the possible seeding process of relativistic electrons and of their energy evolution as a function of time, for a larger sample of clusters. 

\section*{Acknowledgements}
We thank our reviewer, A. Schekochihin, for the helpful and constructive referee report to the first version of this article.
 F.V. acknowledges the financial support by the H2020 initiative, through the ERC StG MAGCOW (n. 714196) and from the Cariplo "BREAKTHRU" funds Rif: 2022-2088 CUP J33C22004310003.  E.M.C. is supported by MIUR grant PRIN 2017 20173ML3WW-001 and Padua University grants DOR2020-2022
\bibliographystyle{aa}
\bibliography{franco3}

\begin{thebibliography}{53}
\expandafter\ifx\csname natexlab\endcsname\relax\def\natexlab#1{#1}\fi

\bibitem[{{Angelinelli} {et~al.}(2020){Angelinelli}, {Vazza}, {Giocoli}, {Ettori}, {Jones}, {Brunetti}, {Br{\"u}ggen}, \& {Eckert}}]{2020MNRAS.495..864A}
{Angelinelli}, M., {Vazza}, F., {Giocoli}, C., {et~al.} 2020, \mnras, 495, 864

\bibitem[{{Beresnyak} \& {Miniati}(2016{\natexlab{a}})}]{2016ApJ...817..127B}
{Beresnyak}, A. \& {Miniati}, F. 2016{\natexlab{a}}, \apj, 817, 127

\bibitem[{{Beresnyak} \& {Miniati}(2016{\natexlab{b}})}]{bm16}
{Beresnyak}, A. \& {Miniati}, F. 2016{\natexlab{b}}, \apj, 817, 127

\bibitem[{{Bonafede} {et~al.}(2021){Bonafede}, {Brunetti}, {Vazza}, {Simionescu}, {Giovannini}, {Bonnassieux}, {Shimwell}, {Br{\"u}ggen}, {van Weeren}, {Botteon}, {Brienza}, {Cassano}, {Drabent}, {Feretti}, {de Gasperin}, {Gastaldello}, {di Gennaro}, {Rossetti}, {Rottgering}, {Stuardi}, \& {Venturi}}]{bonafede21}
{Bonafede}, A., {Brunetti}, G., {Vazza}, F., {et~al.} 2021, \apj, 907, 32

\bibitem[{{B{\"o}ss} {et~al.}(2023){B{\"o}ss}, {Steinwandel}, {Dolag}, \& {Lesch}}]{2023MNRAS.519..548B}
{B{\"o}ss}, L.~M., {Steinwandel}, U.~P., {Dolag}, K., \& {Lesch}, H. 2023, \mnras, 519, 548

\bibitem[{{Botteon} {et~al.}(2020){Botteon}, {van Weeren}, {Brunetti}, {de Gasperin}, {Intema}, {Osinga}, {Di Gennaro}, {Shimwell}, {Bonafede}, {Br{\"u}ggen}, {Cassano}, {Cuciti}, {Dallacasa}, {Gastaldello}, {Mandal}, {Rossetti}, \& {R{\"o}ttgering}}]{2020MNRAS.499L..11B}
{Botteon}, A., {van Weeren}, R.~J., {Brunetti}, G., {et~al.} 2020, \mnras, 499, L11

\bibitem[{{Botteon} {et~al.}(2022){Botteon}, {van Weeren}, {Brunetti}, {Vazza}, {Shimwell}, {Br{\"u}ggen}, {R{\"o}ttgering}, {de Gasperin}, {Akamatsu}, {Bonafede}, {Cassano}, {Cuciti}, {Dallacasa}, {Di Gennaro}, \& {Gastaldello}}]{2022SciA....8.7623B}
{Botteon}, A., {van Weeren}, R.~J., {Brunetti}, G., {et~al.} 2022, Science Advances, 8, eabq7623

\bibitem[{{Brunetti}(2016)}]{2016PPCF...58a4011B}
{Brunetti}, G. 2016, Plasma Physics and Controlled Fusion, 58, 014011

\bibitem[{{Brunetti} \& {Jones}(2014)}]{bj14}
{Brunetti}, G. \& {Jones}, T.~W. 2014, International Journal of Modern Physics D, 23, 1430007

\bibitem[{{Brunetti} \& {Lazarian}(2007)}]{bl07}
{Brunetti}, G. \& {Lazarian}, A. 2007, \mnras, 1371

\bibitem[{{Brunetti} \& {Lazarian}(2016)}]{2016MNRAS.458.2584B}
{Brunetti}, G. \& {Lazarian}, A. 2016, \mnras, 458, 2584

\bibitem[{{Brunetti} {et~al.}(2001){Brunetti}, {Setti}, {Feretti}, \& {Giovannini}}]{gb01}
{Brunetti}, G., {Setti}, G., {Feretti}, L., \& {Giovannini}, G. 2001, \mnras, 320, 365

\bibitem[{{Brunetti} \& {Vazza}(2020)}]{bv20}
{Brunetti}, G. \& {Vazza}, F. 2020, \prl, 124, 051101

\bibitem[{{Bryan} {et~al.}(2014){Bryan}, {Norman}, {O'Shea}, {Abel}, {Wise}, {Turk}, {Reynolds}, {Collins}, {Wang}, {Skillman}, {Smith}, {Harkness}, {Bordner}, {Kim}, {Kuhlen}, {Xu}, {Goldbaum}, {Hummels}, {Kritsuk}, {Tasker}, {Skory}, {Simpson}, {Hahn}, {Oishi}, {So}, {Zhao}, {Cen}, {Li}, \& {Enzo Collaboration}}]{enzo14}
{Bryan}, G.~L., {Norman}, M.~L., {O'Shea}, B.~W., {et~al.} 2014, \apjs, 211, 19

\bibitem[{{Cassano} \& {Brunetti}(2005)}]{cassano05}
{Cassano}, R. \& {Brunetti}, G. 2005, \mnras, 357, 1313

\bibitem[{{Cassano} {et~al.}(2010){Cassano}, {Brunetti}, {R{\"o}ttgering}, \& {Br{\"u}ggen}}]{2010A&A...509A..68C}
{Cassano}, R., {Brunetti}, G., {R{\"o}ttgering}, H.~J.~A., \& {Br{\"u}ggen}, M. 2010, \aap, 509, A68

\bibitem[{{Cho} \& {Lazarian}(2006)}]{2006ApJ...638..811C}
{Cho}, J. \& {Lazarian}, A. 2006, \apj, 638, 811

\bibitem[{{Cuciti} {et~al.}(2022){Cuciti}, {de Gasperin}, {Br{\"u}ggen}, {Vazza}, {Brunetti}, {Shimwell}, {Edler}, {van Weeren}, {Botteon}, {Cassano}, {Di Gennaro}, {Gastaldello}, {Drabent}, {R{\"o}ttgering}, \& {Tasse}}]{2022Natur.609..911C}
{Cuciti}, V., {de Gasperin}, F., {Br{\"u}ggen}, M., {et~al.} 2022, \nat, 609, 911

\bibitem[{{Dom{\'\i}nguez-Fern{\'a}ndez} {et~al.}(2019){Dom{\'\i}nguez-Fern{\'a}ndez}, {Vazza}, {Br{\"u}ggen}, \& {Brunetti}}]{dom19}
{Dom{\'\i}nguez-Fern{\'a}ndez}, P., {Vazza}, F., {Br{\"u}ggen}, M., \& {Brunetti}, G. 2019, \mnras, 486, 623

\bibitem[{{Donnert} \& {Brunetti}(2014)}]{2014MNRAS.443.3564D}
{Donnert}, J. \& {Brunetti}, G. 2014, \mnras, 443, 3564

\bibitem[{{Duchesne} {et~al.}(2021){Duchesne}, {Johnston-Hollitt}, \& {Bartalucci}}]{2021PASA...38...53D}
{Duchesne}, S.~W., {Johnston-Hollitt}, M., \& {Bartalucci}, I. 2021, \pasa, 38, e053

\bibitem[{{Dupourqu{\'e}} {et~al.}(2023){Dupourqu{\'e}}, {Clerc}, {Pointecouteau}, {Eckert}, {Ettori}, \& {Vazza}}]{2023arXiv230315102D}
{Dupourqu{\'e}}, S., {Clerc}, N., {Pointecouteau}, E., {et~al.} 2023, arXiv e-prints, arXiv:2303.15102

\bibitem[{{Fujita} {et~al.}(2003){Fujita}, {Takizawa}, \& {Sarazin}}]{2003ApJ...584..190F}
{Fujita}, Y., {Takizawa}, M., \& {Sarazin}, C.~L. 2003, \apj, 584, 190

\bibitem[{{Girichidis} {et~al.}(2020){Girichidis}, {Pfrommer}, {Hanasz}, \& {Naab}}]{2020MNRAS.491..993G}
{Girichidis}, P., {Pfrommer}, C., {Hanasz}, M., \& {Naab}, T. 2020, \mnras, 491, 993

\bibitem[{{Govoni} {et~al.}(2013){Govoni}, {Murgia}, {Xu}, {Li}, {Norman}, {Feretti}, {Giovannini}, \& {Vacca}}]{2013A&A...554A.102G}
{Govoni}, F., {Murgia}, M., {Xu}, H., {et~al.} 2013, \aap, 554, A102

\bibitem[{{Govoni} {et~al.}(2019){Govoni}, {Orr{\`u}}, {Bonafede}, {Iacobelli}, {Paladino}, {Vazza}, {Murgia}, {Vacca}, {Giovannini}, {Feretti}, {Loi}, {Bernardi}, {Ferrari}, {Pizzo}, {Gheller}, {Manti}, {Br{\"u}ggen}, {Brunetti}, {Cassano}, {de Gasperin}, {En{\ss}lin}, {Hoeft}, {Horellou}, {Junklewitz}, {R{\"o}ttgering}, {Scaife}, {Shimwell}, {van Weeren}, \& {Wise}}]{2019Sci...364..981G}
{Govoni}, F., {Orr{\`u}}, E., {Bonafede}, A., {et~al.} 2019, Science, 364, 981

\bibitem[{{Hoeft} {et~al.}(2021){Hoeft}, {Dumba}, {Drabent}, {Rajpurohit}, {Rossetti}, {Nuza}, {van Weeren}, {Meusinger}, {Botteon}, {Brunetti}, {Shimwell}, {Cassano}, {Br{\"u}ggen}, {R{\"o}ttgering}, {Gastaldello}, {Lovisari}, {Yepes}, {Andrade-Santos}, \& {Eckert}}]{2021A&A...654A..68H}
{Hoeft}, M., {Dumba}, C., {Drabent}, A., {et~al.} 2021, \aap, 654, A68

\bibitem[{{Johnston-Hollitt} {et~al.}(2015){Johnston-Hollitt}, {Govoni}, {Beck}, {Dehghan}, {Pratley}, {Akahori}, {Heald}, {Agudo}, {Bonafede}, {Carretti}, {Clarke}, {Colafrancesco}, {Ensslin}, {Feretti}, {Gaensler}, {Haverkorn}, {Mao}, {Oppermann}, {Rudnick}, {Scaife}, {Schnitzeler}, {Stil}, {Taylor}, \& {Vacca}}]{2015aska.confE..92J}
{Johnston-Hollitt}, M., {Govoni}, F., {Beck}, R., {et~al.} 2015, Advancing Astrophysics with the Square Kilometre Array (AASKA14), 92

\bibitem[{{Kale}(2020)}]{2020IAUS..342...37K}
{Kale}, R. 2020, in Perseus in Sicily: From Black Hole to Cluster Outskirts, ed. K.~{Asada}, E.~{de Gouveia Dal Pino}, M.~{Giroletti}, H.~{Nagai}, \& R.~{Nemmen}, Vol. 342, 37--43

\bibitem[{{Knowles} {et~al.}(2022){Knowles}, {Cotton}, {Rudnick}, {Camilo}, {Goedhart}, {Deane}, {Ramatsoku}, {Bietenholz}, {Br{\"u}ggen}, {Button}, {Chen}, {Chibueze}, {Clarke}, {de Gasperin}, {Ianjamasimanana}, {J{\'o}zsa}, {Hilton}, {Kesebonye}, {Kolokythas}, {Kraan-Korteweg}, {Lawrie}, {Lochner}, {Loubser}, {Marchegiani}, {Mhlahlo}, {Moodley}, {Murphy}, {Namumba}, {Oozeer}, {Parekh}, {Pillay}, {Passmoor}, {Ramaila}, {Ranchod}, {Retana-Montenegro}, {Sebokolodi}, {Sikhosana}, {Smirnov}, {Thorat}, {Venturi}, {Abbott}, {Adam}, {Adams}, {Aldera}, {Bauermeister}, {Bennett}, {Bode}, {Botha}, {Botha}, {Brederode}, {Buchner}, {Burger}, {Cheetham}, {de Villiers}, {Dikgale-Mahlakoana}, {du Toit}, {Esterhuyse}, {Fadana}, {Fanaroff}, {Fataar}, {Foley}, {Fourie}, {Frank}, {Gamatham}, {Gatsi}, {Geyer}, {Gouws}, {Gumede}, {Heywood}, {Hlakola}, {Hokwana}, {Hoosen}, {Horn}, {Horrell}, {Hugo}, {Isaacson}, {Jonas}, {Jordaan}, {Joubert}, {Julie}, {Kapp}, {Kasper}, {Kenyon}, {Kotz{\'e}}, {Kotze}, {Kriek}, {Kriel}, {Krishnan},
  {Kusel}, {Legodi}, {Lehmensiek}, {Liebenberg}, {Lord}, {Lunsky}, {Madisa}, {Magnus}, {Main}, {Makhaba}, {Makhathini}, {Malan}, {Manley}, {Marais}, {Maree}, {Martens}, {Mauch}, {McAlpine}, {Merry}, {Millenaar}, {Mokone}, {Monama}, {Mphego}, {New}, {Ngcebetsha}, {Ngoasheng}, {Ockards}, {Otto}, {Patel}, {Peens-Hough}, {Perkins}, {Ramanujam}, {Ramudzuli}, {Ratcliffe}, {Renil}, {Robyntjies}, {Rust}, {Salie}, {Sambu}, {Schollar}, {Schwardt}, {Schwartz}, {Serylak}, {Siebrits}, {Sirothia}, {Slabber}, {Sofeya}, {Taljaard}, {Tasse}, {Tiplady}, {Toruvanda}, {Twum}, {van Balla}, {van der Byl}, {van der Merwe}, {van Dyk}, {Van Tonder}, {Van Wyk}, {Venter}, {Venter}, {Welz}, {Williams}, \& {Xaia}}]{2022A&A...657A..56K}
{Knowles}, K., {Cotton}, W.~D., {Rudnick}, L., {et~al.} 2022, \aap, 657, A56

\bibitem[{{Kunz} {et~al.}(2022){Kunz}, {Jones}, \& {Zhuravleva}}]{2022hxga.book...56K}
{Kunz}, M.~W., {Jones}, T.~W., \& {Zhuravleva}, I. 2022, in Handbook of X-ray and Gamma-ray Astrophysics. Edited by Cosimo Bambi and Andrea Santangelo, 56

\bibitem[{{Lau} {et~al.}(2009){Lau}, {Kravtsov}, \& {Nagai}}]{lau09}
{Lau}, E.~T., {Kravtsov}, A.~V., \& {Nagai}, D. 2009, \apj, 705, 1129

\bibitem[{{Lemoine} \& {Malkov}(2020)}]{2020MNRAS.499.4972L}
{Lemoine}, M. \& {Malkov}, M.~A. 2020, \mnras, 499, 4972

\bibitem[{{Longair}(2011)}]{longair}
{Longair}, M.~S. 2011, {High Energy Astrophysics}

\bibitem[{{Marcowith} {et~al.}(2020){Marcowith}, {Ferrand}, {Grech}, {Meliani}, {Plotnikov}, \& {Walder}}]{2020LRCA....6....1M}
{Marcowith}, A., {Ferrand}, G., {Grech}, M., {et~al.} 2020, Living Reviews in Computational Astrophysics, 6, 1

\bibitem[{{Nelson} {et~al.}(2014){Nelson}, {Lau}, {Nagai}, {Rudd}, \& {Yu}}]{2014ApJ...782..107N}
{Nelson}, K., {Lau}, E.~T., {Nagai}, D., {Rudd}, D.~H., \& {Yu}, L. 2014, \apj, 782, 107

\bibitem[{{Osinga} {et~al.}(2021){Osinga}, {van Weeren}, {Boxelaar}, {Brunetti}, {Botteon}, {Br{\"u}ggen}, {Shimwell}, {Bonafede}, {Best}, {Bonato}, {Cassano}, {Gastaldello}, {di Gennaro}, {Hardcastle}, {Mandal}, {Rossetti}, {R{\"o}ttgering}, {Sabater}, \& {Tasse}}]{2021A&A...648A..11O}
{Osinga}, E., {van Weeren}, R.~J., {Boxelaar}, J.~M., {et~al.} 2021, \aap, 648, A11

\bibitem[{{Petrosian}(2012)}]{2012SSRv..173..535P}
{Petrosian}, V. 2012, \ssr, 173, 535

\bibitem[{{Pinzke} {et~al.}(2017){Pinzke}, {Oh}, \& {Pfrommer}}]{2017MNRAS.465.4800P}
{Pinzke}, A., {Oh}, S.~P., \& {Pfrommer}, C. 2017, \mnras, 465, 4800

\bibitem[{{Rincon} {et~al.}(2016){Rincon}, {Califano}, {Schekochihin}, \& {Valentini}}]{2016PNAS..113.3950R}
{Rincon}, F., {Califano}, F., {Schekochihin}, A.~A., \& {Valentini}, F. 2016, Proceedings of the National Academy of Science, 113, 3950

\bibitem[{{Schekochihin} {et~al.}(2004){Schekochihin}, {Cowley}, {Taylor}, {Maron}, \& {McWilliams}}]{2004ApJ...612..276S}
{Schekochihin}, A.~A., {Cowley}, S.~C., {Taylor}, S.~F., {Maron}, J.~L., \& {McWilliams}, J.~C. 2004, \apj, 612, 276

\bibitem[{{Schlickeiser} {et~al.}(1987){Schlickeiser}, {Sievers}, \& {Thiemann}}]{1987A&A...182...21S}
{Schlickeiser}, R., {Sievers}, A., \& {Thiemann}, H. 1987, \aap, 182, 21

\bibitem[{{Simonte} {et~al.}(2022){Simonte}, {Vazza}, {Brighenti}, {Br{\"u}ggen}, {Jones}, \& {Angelinelli}}]{2022A&A...658A.149S}
{Simonte}, M., {Vazza}, F., {Brighenti}, F., {et~al.} 2022, \aap, 658, A149

\bibitem[{Sironi {et~al.}(2023)Sironi, Comisso, \& Golant}]{PhysRevLett.131.055201}
Sironi, L., Comisso, L., \& Golant, R. 2023, Phys. Rev. Lett., 131, 055201

\bibitem[{{St-Onge} \& {Kunz}(2018)}]{2018ApJ...863L..25S}
{St-Onge}, D.~A. \& {Kunz}, M.~W. 2018, \apjl, 863, L25

\bibitem[{{van Haarlem} {et~al.}(2013){van Haarlem}, {Wise}, {Gunst}, {Heald}, {McKean}, {Hessels}, {de Bruyn}, {Nijboer}, {Swinbank}, {Fallows}, {Brentjens}, {Nelles}, {Beck}, {Falcke}, {Fender}, {H{\"o}randel}, {Koopmans}, {Mann}, {Miley}, {R{\"o}ttgering}, {Stappers}, {Wijers}, {Zaroubi}, {van den Akker}, {Alexov}, {Anderson}, {Anderson}, {van Ardenne}, {Arts}, {Asgekar}, {Avruch}, {Batejat}, {B{\"a}hren}, {Bell}, {Bell}, {van Bemmel}, {Bennema}, {Bentum}, {Bernardi}, {Best}, {B{\^\i}rzan}, {Bonafede}, {Boonstra}, {Braun}, {Bregman}, {Breitling}, {van de Brink}, {Broderick}, {Broekema}, {Brouw}, {Br{\"u}ggen}, {Butcher}, {van Cappellen}, {Ciardi}, {Coenen}, {Conway}, {Coolen}, {Corstanje}, {Damstra}, {Davies}, {Deller}, {Dettmar}, {van Diepen}, {Dijkstra}, {Donker}, {Doorduin}, {Dromer}, {Drost}, {van Duin}, {Eisl{\"o}ffel}, {van Enst}, {Ferrari}, {Frieswijk}, {Gankema}, {Garrett}, {de Gasperin}, {Gerbers}, {de Geus}, {Grie{\ss}meier}, {Grit}, {Gruppen}, {Hamaker}, {Hassall}, {Hoeft}, {Holties},
  {Horneffer}, {van der Horst}, {van Houwelingen}, {Huijgen}, {Iacobelli}, {Intema}, {Jackson}, {Jelic}, {de Jong}, {Juette}, {Kant}, {Karastergiou}, {Koers}, {Kollen}, {Kondratiev}, {Kooistra}, {Koopman}, {Koster}, {Kuniyoshi}, {Kramer}, {Kuper}, {Lambropoulos}, {Law}, {van Leeuwen}, {Lemaitre}, {Loose}, {Maat}, {Macario}, {Markoff}, {Masters}, {McFadden}, {McKay-Bukowski}, {Meijering}, {Meulman}, {Mevius}, {Middelberg}, {Millenaar}, {Miller-Jones}, {Mohan}, {Mol}, {Morawietz}, {Morganti}, {Mulcahy}, {Mulder}, {Munk}, {Nieuwenhuis}, {van Nieuwpoort}, {Noordam}, {Norden}, {Noutsos}, {Offringa}, {Olofsson}, {Omar}, {Orr{\'u}}, {Overeem}, {Paas}, {Pandey-Pommier}, {Pandey}, {Pizzo}, {Polatidis}, {Rafferty}, {Rawlings}, {Reich}, {de Reijer}, {Reitsma}, {Renting}, {Riemers}, {Rol}, {Romein}, {Roosjen}, {Ruiter}, {Scaife}, {van der Schaaf}, {Scheers}, {Schellart}, {Schoenmakers}, {Schoonderbeek}, {Serylak}, {Shulevski}, {Sluman}, {Smirnov}, {Sobey}, {Spreeuw}, {Steinmetz}, {Sterks}, {Stiepel}, {Stuurwold},
  {Tagger}, {Tang}, {Tasse}, {Thomas}, {Thoudam}, {Toribio}, {van der Tol}, {Usov}, {van Veelen}, {van der Veen}, {ter Veen}, {Verbiest}, {Vermeulen}, {Vermaas}, {Vocks}, {Vogt}, {de Vos}, {van der Wal}, {van Weeren}, {Weggemans}, {Weltevrede}, {White}, {Wijnholds}, {Wilhelmsson}, {Wucknitz}, {Yatawatta}, {Zarka}, {Zensus}, \& {van Zwieten}}]{2013A&A...556A...2V}
{van Haarlem}, M.~P., {Wise}, M.~W., {Gunst}, A.~W., {et~al.} 2013, \aap, 556, A2

\bibitem[{{van Weeren} {et~al.}(2019){van Weeren}, {de Gasperin}, {Akamatsu}, {Br{\"u}ggen}, {Feretti}, {Kang}, {Stroe}, \& {Zandanel}}]{2019SSRv..215...16V}
{van Weeren}, R.~J., {de Gasperin}, F., {Akamatsu}, H., {et~al.} 2019, \ssr, 215, 16

\bibitem[{{Vazza} {et~al.}(2011){Vazza}, {Brunetti}, {Gheller}, {Brunino}, \& {Br{\"u}ggen}}]{va11turbo}
{Vazza}, F., {Brunetti}, G., {Gheller}, C., {Brunino}, R., \& {Br{\"u}ggen}, M. 2011, \aap, 529, A17+

\bibitem[{{Vazza} {et~al.}(2017){Vazza}, {Jones}, {Br{\"u}ggen}, {Brunetti}, {Gheller}, {Porter}, \& {Ryu}}]{va17turb}
{Vazza}, F., {Jones}, T.~W., {Br{\"u}ggen}, M., {et~al.} 2017, \mnras, 464, 210

\bibitem[{{Vazza} {et~al.}(2021){Vazza}, {Wittor}, {Brunetti}, \& {Br{\"u}ggen}}]{va21jets}
{Vazza}, F., {Wittor}, D., {Brunetti}, G., \& {Br{\"u}ggen}, M. 2021, \aap, 653, A23

\bibitem[{{Vazza} {et~al.}(2023){Vazza}, {Wittor}, {Di Federico}, {Br{\"u}ggen}, {Brienza}, {Brunetti}, {Brighenti}, \& {Pasini}}]{va23a}
{Vazza}, F., {Wittor}, D., {Di Federico}, L., {et~al.} 2023, \aap, 669, A50

\bibitem[{{Wittor} {et~al.}(2017){Wittor}, {Vazza}, \& {Br{\"u}ggen}}]{wi17}
{Wittor}, D., {Vazza}, F., \& {Br{\"u}ggen}, M. 2017, \mnras, 464, 4448

\bibitem[{Zhou {et~al.}(2023)Zhou, Zhdankin, Kunz, Loureiro, \& Uzdensky}]{zhou2023magnetogenesis}
Zhou, M., Zhdankin, V., Kunz, M.~W., Loureiro, N.~F., \& Uzdensky, D.~A. 2023, Magnetogenesis in a collisionless plasma: from Weibel instability to turbulent dynamo

\end{thebibliography}

\end{document}